\documentclass[amstex,thmsa, 12pt,sw20aip]{article}
\usepackage{amssymb}
\usepackage{amsfonts}
\usepackage{amsmath}
\usepackage{graphicx}
\begin{document}

\begin{center}
{\large Hot Brownian Carriers in the Langevin Picture:}

{\large \ Application to a simple model for the Gunn Effect in GaAs. }

{\large \ \vspace{0.3cm}}

F. C. Bauke$^{a)}$ and R. E. Lagos$^{b)}$

$^{a)}$ Instituto de F\'{\i}sica \textit{Gleb Wataghin }UNICAMP\
(Universidade Estadual de Campinas) \\[0pt]
CP $6165$, $13083-970$ Campinas, SP Brazil.\\[0pt]

$^{b)}$ Departamento de F\'{\i}sica, IGCE, UNESP (Universidade Estadual
Paulista)\\[0pt]
CP $178$, $13500$-$970$ Rio Claro SP Brazil.\\[0pt]

\vspace{1cm}
\end{center}

We consider a charged Brownian gas under the influence of external, static
and uniform electric and magnetic fields, immersed in a uniform bath
temperature. We obtain the solution for the associated Langevin equation,
and thereafter the evolution of the nonequilibrium temperature towards a
nonequilibrium (hot) steady state. We apply our results to a simple yet
relevant Brownian model for carrier transport in GaAs. We obtain a negative
differential conductivity regime (Gunn effect) and discuss and compare our
results with experimental results.

\bigskip

\noindent Pacs {05.20.Dd, 05.40.-a, 05.40.Jc, 05.70.Ln, 51.10+y, 66.10.Cd,
82.40.-g}


\noindent Corresponding author: R. E. Lagos, monaco@rc.unesp.br

\vspace{0.5cm}\newpage

\begin{center}
{\large 1: Introduction }

\bigskip
\end{center}

The ubiquitous Brownian motion remains an outstanding paradigm in modern
physics. Some representative, but by no means an exhaustive list of general
references ("founding papers", reviews and applications) are presented in
[1-19]. Here we present the Langevin formulation for a Brownian carrier in
uniform and static external fields. Some recent work on charged Brownian
particles is referenced in [20-47]. In our previous work on this matter, our
approach hinged on the resolution of Kramers and/or Smoluchowski equations
[24,26,28,45,46,47], and recently we began to tackle Langevin%
\'{}%
s formulation of this problem [47]. \ Here we explore the latter, in order
to study the relaxation of the Brownian carrier towards a steady state,
given electrical and magnetic external static and uniform fields. \ In
section 2 we present the solution of Langevin%
\'{}%
s equation \ including the above mentioned fields. In section 3 we present
our results for the nonequilbrium temperature relaxation to the "hot" steady
state temperature (as modified by the electric and magnetic fields). The
computed final "hot" regime temperature is compared to long time existing
results ([48] with no magnetic field present) \ and with our previous
results, with the magnetic field contribution, via Kramers and Smoluchowski
equations [26,28,45]. In section 4 we present an application, namely a
simple yet relevant Brownian model\ (with no adjustable parameters) for GaAs
carrier mobility [49-56]. The multivalley band structure, and the "hot"
carrier \ steady state temperature obtained in the previous section are the
essential ingredients for the appearance of a negative differential
conductivity regime, in good quantitative agreement with well known
experimental results. Furthermore our model incorporates the magnetic field
contribution hitherto not considered. Finally , in section 6 we present our
concluding remarks and outline further work. \ 

\bigskip

\begin{center}
{\large 2: Langevin equation for a Brownian charged particle }
\end{center}

We briefly present the Langevin formalism for a free charged Brownian 
particle [10-12,18,19], with mass $m$, and charge $q$ immersed in an
homogeneous thermal reservoir at temperature $T_{R}$. It is essentially
Newton$^\prime$s equation for the particle with two contributing forces: the first, a
systematic dissipative force Stokes like (linear in the particle%
\'{}%
s velocity) and the second a rapidly fluctuating random force,

\begin{equation}
m\frac{d\mathbf{v}}{dt}=\mathbf{F}_{\text{S}}+\mathbf{F}^{\text{r}}=-\gamma 
\mathbf{v+F}^{\text{r}}(t)\hspace{1cm}\tau =\frac{m}{\gamma }\hspace{1cm}
\end{equation}

The formal solution is

\begin{equation}
\mathbf{v}(t)=\exp \left( -\frac{t}{\tau }\right) \mathbf{v}^{0}+\frac{1}{m}%
\int_{0}^{t}dt_{1}\exp \left( \frac{t_{1}-t}{\tau }\right) \mathbf{F}^{\text{%
r}}(t_{1})
\end{equation}%
with initial condition $\mathbf{v}^{0}=\mathbf{v}(0)$ and \ $\tau $ the
collision time. The random force has solely statistical properties: zero
average and white noise correlations , given by the averages

\begin{equation}
\left\langle \mathbf{F}^{R}(t)\right\rangle =\mathbf{0\hspace{1cm}}%
\left\langle F_{i}^{\text{r}}(t_{1})F_{j}^{\text{r}}(t_{2})\right\rangle =2%
\frac{m}{\tau }k_{B}T_{R}\delta _{ij}\delta (t_{1}-t_{2})
\end{equation}

\noindent where the correlation strength is such that the asymptotic \
average kinetic energy \ satisfies the equipartition theorem, in thermal
equilibrium with the thermal reservoir(fluctuation dissipation theorem), and
given by

\begin{equation}
\frac{1}{2}m\left\langle \mathbf{v}^{2}(t\rightarrow \infty )\right\rangle =%
\frac{3}{2}k_{B}T_{R}=\frac{1}{2}mV_{T}^{2}
\end{equation}

\noindent

Following [28,45,46,47] (and with a slightly different notation) we now
consider the Brownian carrier (charged particle) under the influence of
homogeneous external, time independent, electric and magnetic fields; the
electric contribution is given by $\mathbf{F}_{\text{elec}}=q\mathbf{E}$ and
the magnetic contribution (Lorentz%
\'{}%
s velocity dependent force) $\mathbf{F}_{\text{mag}}\mathbf{=}\frac{1}{c}q%
\mathbf{v}\times \mathbf{B}$. The total external force is given by

\begin{equation}
\mathbf{F}(\mathbf{v})\mathbf{=\mathbf{\mathbf{F}}}_{\text{elec}}\mathbf{%
\mathbf{+\mathbf{F}}}_{\text{mag}}\mathbf{\mathbf{(v)}=}q\mathbf{E-}m\mathbf{%
\omega \times v\hspace{1cm}\omega }=\frac{q}{mc}\mathbf{B}
\end{equation}

Let us define a tensorial Stokes force \ by adding the Lorentz contribution
to the usual Stokes force, as

\begin{equation}
\mathbf{F}_{\text{TS}}\mathbf{=-}\gamma \mathbf{v-}m\mathbf{\omega \times v=-%
}m\mathbf{\Lambda }^{-1}\mathbf{v\hspace{1cm}}
\end{equation}%
where $\mathbf{\omega }$ is the usual cyclotron frequency, the magneto
mobility tensor is $\mathbf{M=}m^{-1}\mathbf{\Lambda }$\ \ with $\mathbf{%
\Lambda }$\ a tensorial collision time, that can be cast into the form (when
operating over an arbitrary vector $\mathbf{V}$\textbf{)}

\bigskip

\begin{equation}
\mathbf{\Lambda (}\tau \mathbf{,\omega )V}=\tau \frac{\mathbf{V+}\tau 
\mathbf{V}\times \mathbf{\omega +}\tau ^{2}\mathbf{\omega }\left( \mathbf{%
\omega \cdot V}\right) }{1+\tau ^{2}\mathbf{\omega }^{2}}
\end{equation}

\bigskip

In particular notice the familiar form for the case $\mathbf{B=}B\widehat{z} 
$

\begin{equation}
\mathbf{\Lambda (}\tau \mathbf{,\omega )=}\frac{\tau }{1+\tau ^{2}\omega ^{2}%
}\left( 
\begin{array}{ccc}
1 & \tau \omega & 0 \\ 
-\tau \omega & 1 & 0 \\ 
0 & 0 & 1+\tau ^{2}\omega ^{2}%
\end{array}%
\right)
\end{equation}

By defining such a tensorial Stokes force, Langevin's equation now reads

\begin{equation}
m\frac{d\mathbf{v}}{dt}=\mathbf{-}m\mathbf{\Lambda }^{-1}\mathbf{v+}q\mathbf{%
E+F}^{\text{r}}(t)
\end{equation}

with formal solution [57-60]$\hspace{1cm}$

\begin{equation}
\mathbf{v}(t)=\exp \left( -\mathbf{\Lambda }^{-1}t\right) \mathbf{v}^{0}+%
\mathbf{\Lambda }\left( 1-\exp \left( -\mathbf{\Lambda }^{-1}t\right)
\right) \frac{q\mathbf{E}}{m}+\frac{1}{m}\int_{0}^{t}dt_{1}\exp \left( 
\mathbf{\Lambda }^{-1}\mathbf{(}t_{1}-t)\right) \mathbf{F}^{\text{r}}(t_{1})
\end{equation}

Using Cayley-Hamilton theorem, and Putzer [59] and Apostol [60] results,
after a lengthily but straight forward calculation we obtain

\bigskip \bigskip 
\begin{equation}
\exp \left( \mathbf{\Lambda }^{-1}t\right) =a_{0}(t)+a_{1}(t)\mathbf{\Lambda 
}^{-1}+a_{2}(t)\mathbf{\Lambda }^{-2}
\end{equation}%
\begin{equation}
=\exp \left( \frac{t}{\tau }\right) \left( 1+\frac{1}{\omega ^{2}}(\mathbf{%
\Lambda }^{-1}\mathbf{-\tau }^{-1})^{2}(1-\cos \omega t)+\frac{1}{\omega }(%
\mathbf{\Lambda }^{-1}\mathbf{-\tau }^{-1})\sin \omega t\right)
\end{equation}

\newpage

\begin{center}
{\large 3: Evolution of the Effective (nonequilibrium) temperature towards a
hot steady state.}
\end{center}

{\large \ }

Let us define the nonequilibrium effective temperature $T_{\text{ef}}(t)$ as

\begin{equation}
\frac{3}{2}k_{B}T_{\text{ef}}(t)=\frac{1}{2}m\left\langle \mathbf{v}%
^{2}(t)\right\rangle
\end{equation}%
and consider "thermal like" initial velocity conditions,\emph{\ i.e.}
configurational averages $\left\langle {}\right\rangle _{\text{c}}$ over
initial velocities are given by Maxwell like distributions

\begin{equation}
\left\langle \mathbf{v}^{0}\right\rangle _{\text{c}}=\mathbf{0\hspace{1cm}}%
\frac{1}{2}m\left\langle v_{i}^{0}v_{j}^{0}\right\rangle _{\text{c}}=\frac{1%
}{2}k_{B}T_{0}\delta _{ij}
\end{equation}%
where $T_{0}<T_{R}$ \ \ ($T_{0}>T_{R}$) defines \ cold (hot) initial
distributions. Furthermore we define dimensionless electric and magnetic
fields $\mathbf{e}$ and $\mathbf{b}$ as

\begin{equation}
\mathbf{e}^{2}=\frac{\tau ^{2}q^{2}\mathbf{E}^{2}}{3mk_{B}T_{R}}=\left( 
\frac{\mathbf{V}_{E}}{V_{T}}\right) ^{2}\hspace{1cm}\mathbf{V}_{E}=\frac{%
q\tau \mathbf{E}}{m}
\end{equation}

\begin{equation}
\mathbf{b=}\tau \mathbf{\omega \hspace{1cm}eb=}eb\cos \theta \hspace{1cm}%
\omega =\left\vert \mathbf{\omega }\right\vert
\end{equation}%
and $\theta $ the angle between fields. \ Furthermore we introduce a
dimensionless time in collision time units and denoted hereafter as $%
t=t/\tau .$ Finally, \ by considering \ equation (3), some trivial
integration and basic vector and matrix algebra, the expression for the
effective temperature is given by

\begin{equation}
T_{\text{ef}}(t)=T_{0}\exp \left( -2t\right) +T_{R}\left( 1-\exp \left(
-2t\right) \right) +T_{R}\mathbf{e}^{2}\left( \frac{1+b^{2}\cos ^{2}\theta }{%
1+b^{2}}+\Gamma _{1}(t)+\Gamma _{2}(t)\right)
\end{equation}

with

\begin{equation}
\Gamma _{1}(t)=\frac{2\exp \left( -t\right) }{1+b^{2}}\left( \left( \cos bt+%
\frac{2b}{1+b^{2}}\sin bt\right) \sin ^{2}\theta +\cos ^{2}\theta \right)
\end{equation}

\begin{equation}
\Gamma _{2}(t)=\exp \left( -2t\right) \left( 1+b^{2}\cos ^{2}\theta +\frac{4b%
}{\left( 1+b^{2}\right) ^{2}}\cos bt\sin bt\sin ^{2}\theta +\cos ^{2}\theta
\right)
\end{equation}%
\bigskip

The Brownian carrier gas evolves towards a stationary state with a
nonequilibrium temperature greater that the reservoir temperature, thereby
the name hot carrier.

\begin{equation}
\Theta =\frac{T_{\text{ef}}(t\rightarrow \infty )}{T_{R}}=1+\left( \frac{%
1+b^{2}\cos ^{2}\theta }{1+b^{2}}\right) \mathbf{e}^{2}
\end{equation}%
This result includes the magnetic field effect. In the Kramers Smoluchowski
scheme [26,28,45] we derived the expression

\bigskip 
\begin{equation}
\Theta _{\text{K}}=\frac{T_{\text{ef}}^{\text{Kramers}}(t\rightarrow \infty )%
}{T_{R}}=1-\frac{1}{t}+\left( \frac{1+b^{2}\cos ^{2}\theta }{1+b^{2}}\right)
\left( \mathbf{e-}\tau V_{T}\frac{\mathbf{\nabla }n(\mathbf{x})}{n(\mathbf{x}%
)}\right) ^{2}
\end{equation}%
where a long tail ($1/t$ non exponential tail, see also [25])and the spatial
inhomogeneity of the carrier's distribution corrects the electric field
contribution. Both the long tail and the density diffusive effect smears out
at longer times rendering both expressions equivalent. We believe equations
(20) and (21) represent novel results at least in the Brownian context, as
far as the magnetic field contribution is concerned. Shockley obtained an
expression for the hot carrier temperature [48-50] for null magnetic field
and in agreement with our result in equation (20) \ for $b=0$. The absence
of the electric field renders the equilibrium (reservoir) temperature,
regardless of the magnetic field value. For non zero electric field values,
the magnetic field modulates the hot carrier's temperature with maximum
value for parallel fields (independent of the magnetic field value) and
minimum value for perpendicular fields.

Figures 1 to 5 we plot the effective temperature evolution, with unit time
given by the collision time constant $\tau .$ Solid lines represent the zero
magnetic field and dashed lines the magnetic and angle value on display in
each figure. \ The reservoir temperature is fixed at $400^{0}$K and two
values are chosen for $T_{0}$ namely $200^{0}$K (cold) and $600^{0}$K (hot) $%
\ $initial velocity distributions. The cold (hot) cases are the lower
(upper) curves starting at $t=0$. We observe that after a few collision time
units the carrier's temperature reaches the stationary value that decreases
as the magnetic field deviates from the parallel configuration, similarly
for the non parallel case as the magnetic field increases. The transient
effects are damped oscillations, a larger effect for intermediate magnetic
field values. The dimensionless field values where chosen from typical
material parameters for GaAs (see next section), rendering: $e=1$
corresponds to $\ 1000$ volts/cm and $b=1$ corresponds to one Tesla.

\begin{figure}[htb]
\begin{center}
\centerline{\includegraphics[width=17cm,height=12.0cm]{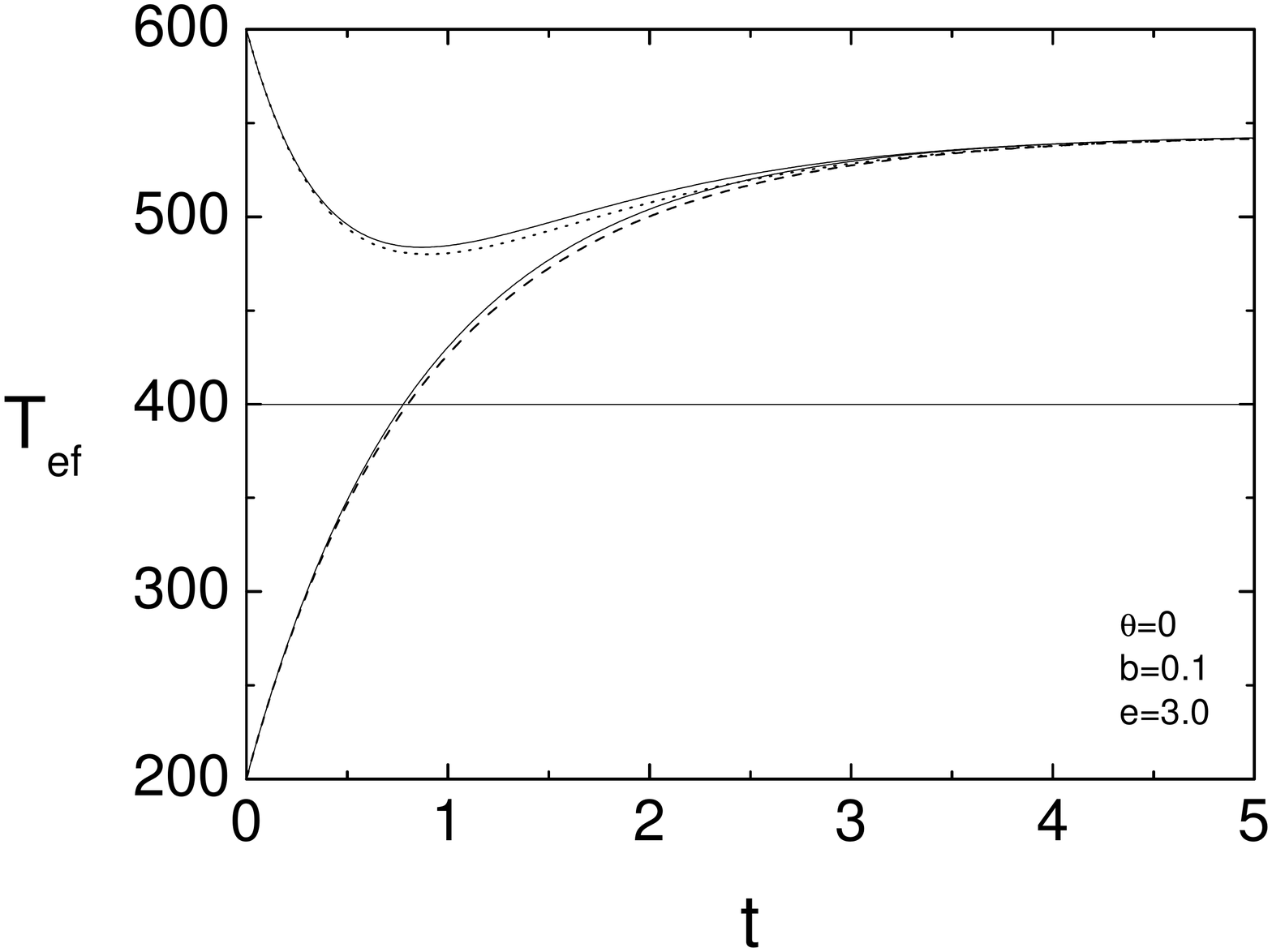}}
\end{center}
\caption{\it{Effective temperature versus
time, electric field $e=3.0$. Solid lines represent the zero magnetic field
case $b=0$, dashed lines the case $b=0.1$, $\ \protect\theta =0.$ Hot
initial condition given by $T_{\text{ef}}(0)=600^{0}K$ and \ cold initial
condition by $T_{\text{ef}}(0)=200^{0}K$. See main text for time and field
units.}}
\label{fig1}
\end{figure}

\begin{center}
\bigskip

\newpage
\end{center}

\begin{figure}[htb]
\begin{center}
\centerline{\includegraphics[width=17cm,height=12.0cm]{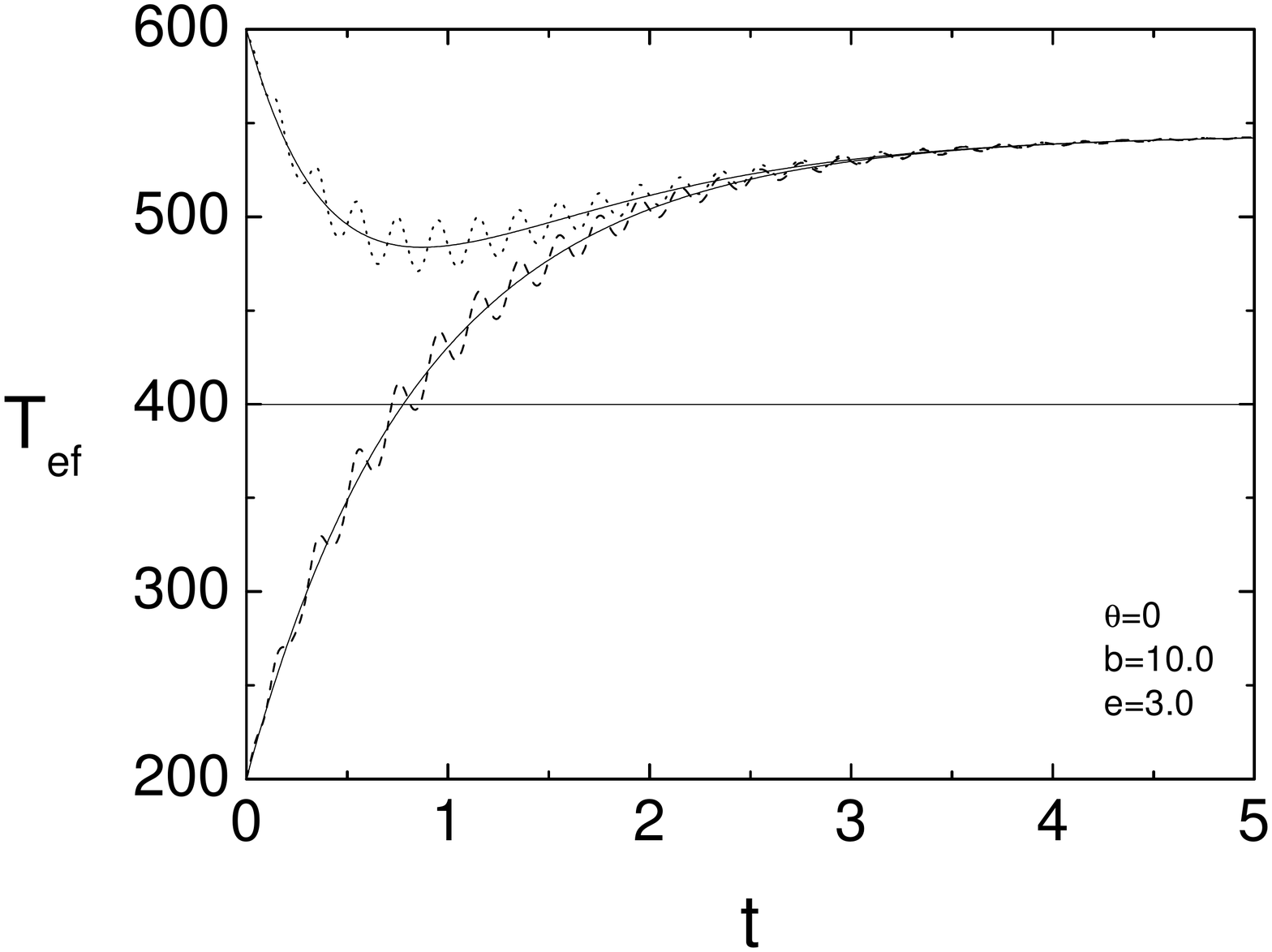}}
\end{center}
\caption{\it{Effective temperature versus
time, electric field $e=3.0$. Solid lines represent the zero magnetic field
case $b=0$, dashed lines the case $b=10.0$, $\ \protect\theta =0.$ Hot
initial condition given by $T_{\text{ef}}(0)=600^{0}K$ and \ cold initial
condition by $T_{\text{ef}}(0)=200^{0}K$. See main text for time and field
units.}}
\label{fig2}
\end{figure}

\begin{center}
\newpage
\end{center}

\begin{figure}[htb]
\begin{center}
\centerline{\includegraphics[width=17cm,height=12.0cm]{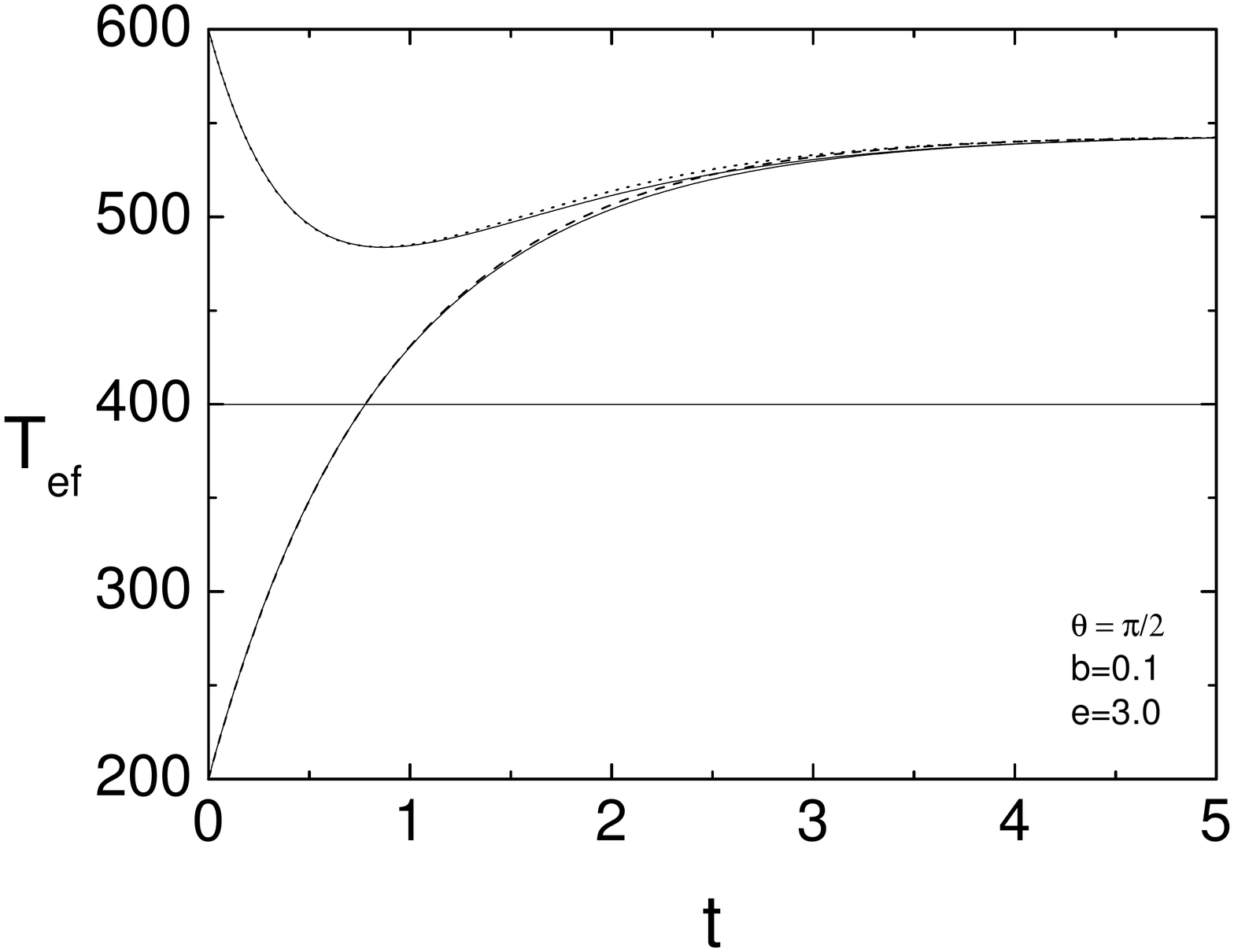}}
\end{center}
\caption{\it{Effective temperature versus
time, electric field $e=3.0$. Solid lines represent the zero magnetic field
case $b=0$, dashed lines the case $b=0.1$, $\ \protect\theta =\protect\pi %
/2. $ Hot initial condition given by $T_{\text{ef}}(0)=600^{0}K$ and \ cold
initial condition by $T_{\text{ef}}(0)=200^{0}K$. See main text for time and
field units.}}
\label{fig3}
\end{figure}

\begin{center}
\bigskip \newpage
\end{center}

\begin{figure}[htb]
\begin{center}
\centerline{\includegraphics[width=17cm,height=12.0cm]{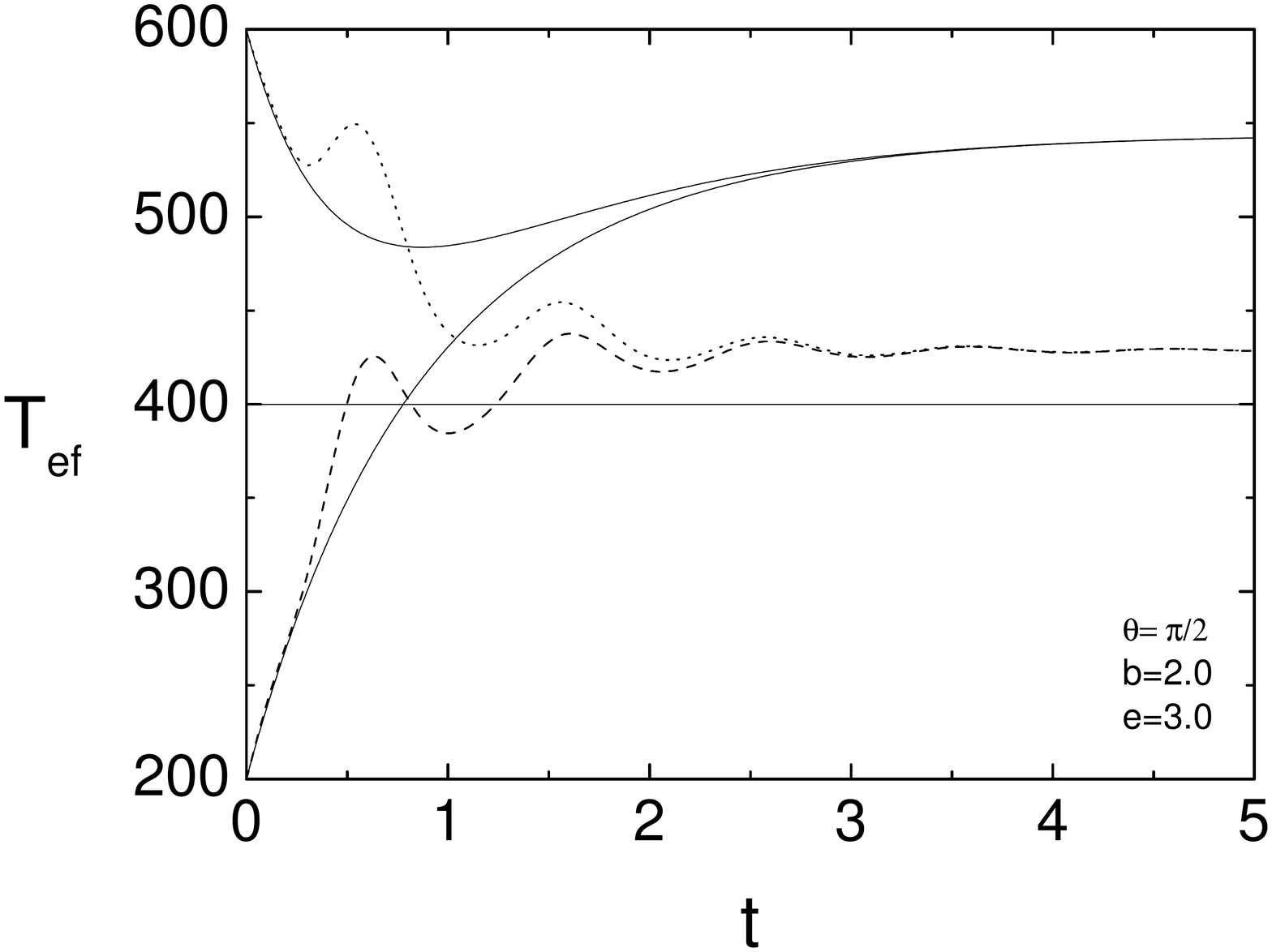}}
\end{center}
\caption{\it{Effective temperature versus
time, electric field $e=3.0$. Solid lines represent the zero magnetic field
case $b=0$, dashed lines the case $b=2.0$, $\ \protect\theta =\protect\pi %
/2. $ Hot initial condition given by $T_{\text{ef}}(0)=600^{0}K$ and \ cold
initial condition by $T_{\text{ef}}(0)=200^{0}K$. See main text for time and
field units.}}
\label{fig4}
\end{figure}

\begin{center}
\bigskip \newpage

\begin{figure}[htb]
\begin{center}
\centerline{\includegraphics[width=17cm,height=12.0cm]{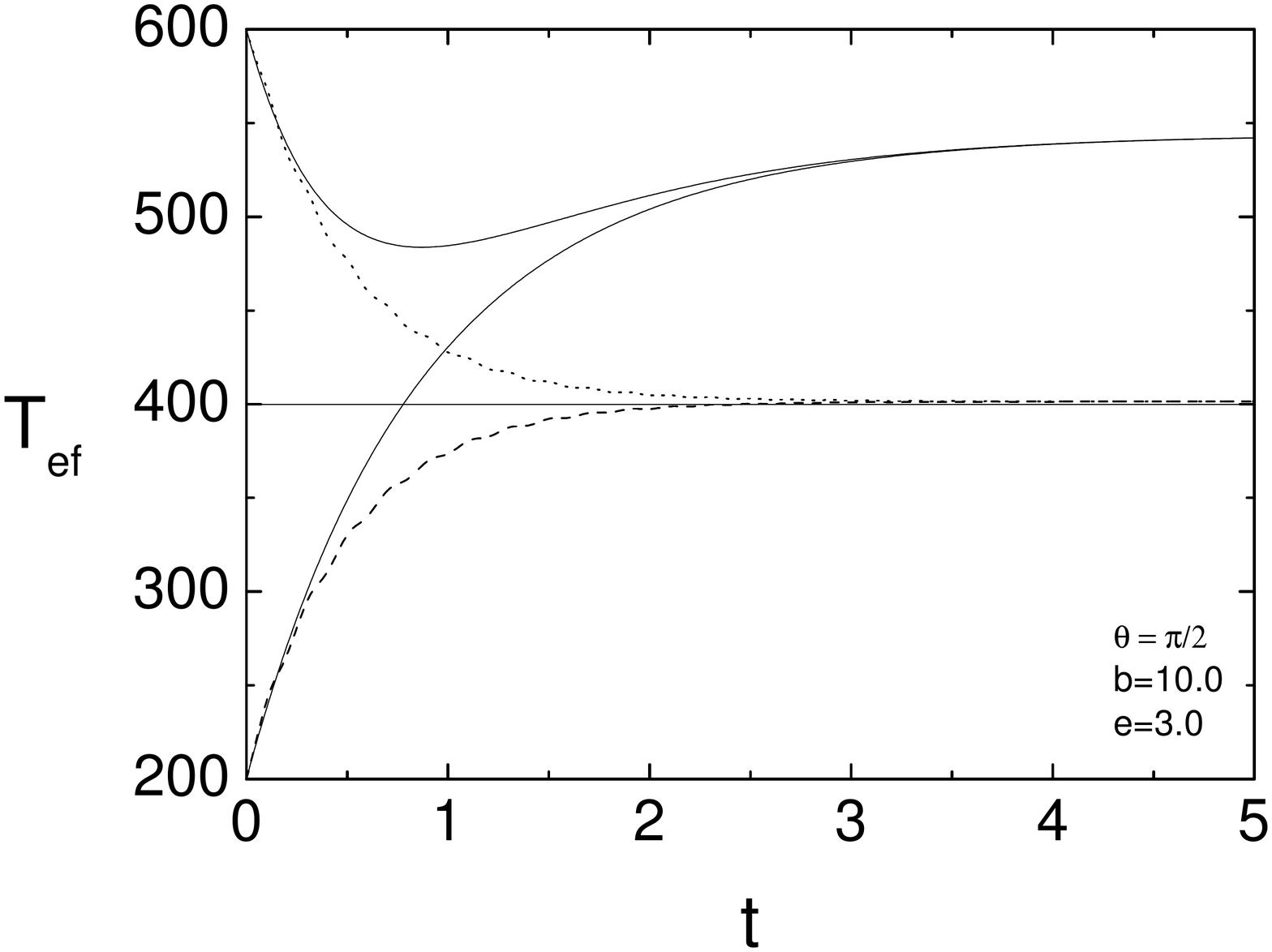}}
\end{center}
\caption{\it{Effective temperature versus
time, electric field $e=3.0$. Solid lines represent the zero magnetic field
case $b=0$, dashed lines the case $b=10.0$, $\ \protect\theta =\protect\pi %
/2.$ Hot initial condition given by $T_{\text{ef}}(0)=600^{0}K$ and \ cold
initial condition by $T_{\text{ef}}(0)=200^{0}K$. See main text for time and
field units.}}
\label{fig5}
\end{figure}

\bigskip \newpage

\bigskip {\large 4: Simple Brownian carrier model for GaAs. Negative
differential conductivity. }
\end{center}

Gallium arsenide (GaAs) is a compound of the elements Gallium and Arsenic.
It is a III/V semiconductor, and is used in the manufacture of devices such
as microwave frequency integrated circuits, monolithic microwave integrated
circuits, infrared light-emitting diodes, laser diodes, solar cells and
optical windows. The band structure consists of a multivalley landscape.
Based on data from references [49-56] we model the electron mobility $\mu $
(proportional to the drift velocity $V_{D}$), combining the high mobility $%
\Gamma $ valley with low mobility L valley (six satellite valleys per $%
\Gamma $ type valley). The carrier mobility is obtained via standard
canonical average restricted to these two type of valleys,%
\begin{equation}
\mu =\frac{\mu _{\Gamma }P_{\Gamma }+\mu _{\text{L}}P_{\text{L}}}{P_{\Gamma
}+P_{\text{L}}}
\end{equation}%
with the usual expressions as given in the literature [49-56]

\ 
\begin{equation}
\mu _{\alpha }=\mu _{0}\sqrt{\frac{1}{m_{\alpha }T}}\hspace{1cm}P_{\alpha
}\sim g_{\alpha }m_{\alpha }^{3/2}\exp \left( -\frac{E_{\alpha }}{k_{B}T}%
\right) \hspace{1cm}\alpha =\Gamma ,\text{L}
\end{equation}%
where $\mu _{0}$ is proportional to a typical mean velocity equation (10)
and with relevant typical parameters given by: $g_{\text{L}}=6g_{\Gamma },$ $%
m_{\text{L}}\simeq 10m_{\Gamma }$ and $\Delta =E_{\text{L}}-E_{\Gamma
}\simeq 0.3$eV. At room temperature 300$^{0}K$ \ \ we have \ $\Delta
/k_{B}T_{R}\simeq 12.$ The drift velocity (mobility times electric field) is
computed in the electric field direction as (in arbitrary units).

\begin{equation}
V_{D}=\frac{\mu }{\mu _{0}}\frac{\mathbf{v}(t\rightarrow \infty )\mathbf{E}}{%
\left\vert \mathbf{E}\right\vert }
\end{equation}

With typical values for GaAs \ we compute the last equation \ using
equations (10) and (22) yielding the material parameter free equations

\bigskip

\begin{equation}
V_{D}=\frac{1}{\sqrt{\Theta }}\left( \frac{1+60\exp \left( -12\Theta
^{-1}\right) }{1+130\exp \left( -12\Theta ^{-1}\right) }\right) \Omega
\left\vert \mathbf{e}\right\vert
\end{equation}

\begin{equation}
\Omega =\frac{1+\mathbf{b}^{2}\cos ^{2}\theta }{1+\mathbf{b}^{2}}
\end{equation}

\begin{equation}
\Theta =1+0.04\Omega \mathbf{e}^{2}
\end{equation}%
notice that for zero gap we obtain a typical Caughey Thomas [56] expression
for the mobility.

\begin{equation}
V_{D}^{\Delta =0}=\frac{0.47\Omega }{\sqrt{1+0.04\Omega \mathbf{e}^{2}}}%
\left\vert \mathbf{e}\right\vert
\end{equation}

Caughey Thomas mobility modeling is a non linear phenomenological fitting
procedure. Our general result is linear in the electric field (in the linear
regime, the current is \ a transport coefficient times the field),
incorporates the magnetic field, and the nonlinear electric field dependence
is due to the non equilibrium temperature intrinsic field dependence. As
mentioned in the previous section, the dimensionless field values where
chosen from typical material parameters for GaAs, rendering: $e=1$ $%
\rightarrow $ $E=1000$ volts/cm and $b=1\rightarrow $ $B=1$Tesla.

In Figures 6 to 9 \ we plot the drift velocity in arbitrary units versus
electric field. The solid upper line is the zero magnetic field case and the
lower solid lines is the corresponding Caughey Thomas case ($\Delta =0$).
The dashed lines are corresponds to the angle and magnetic field values on
display in each figure, the dashed lines maximum decreases as the magnetic
field value increases. For all cases presented we find a region of negative
differential conductivity \ (Gunn effect)

\bigskip 
\begin{equation*}
\frac{dV_{D}(e)}{de}<0
\end{equation*}%
for electric field larger than the critical value $\ e_{c}\sim 4000$
Volts/cm, $\left( \frac{dV_{D}(e_{c})}{de_{c}}=0\right) $ as it is well
known for this compound [49,50,52-54]. From Figures 6-9 and considering
equations (25-27), as the magnetic field value is increased we notice that $%
e_{c}(b)$ increases and $V_{D}(e_{c}(b))$ decreases while the effective
temperature $\Theta (e_{c}(b),b,\theta )$ decreases. This pattern becomes
more pronounced as we move from parallel fields towards perpendicular
fields. The available experimental data for a related compound [55], where
the negative differential conductivity region was probed with a magnetic
field, seems to corroborate our findings, in a very qualitative fashion. In
the higher electric field regions, say $e\gtrsim 10$ where our results
deviates from the experimental data, the effective temperature $\Theta $ is
very large, quite probably other scattering mechanisms should be
incorporated, rendering a more involved temperature dependence for the
mobility (see equation (23)), and more than one collision time constant
should be considered. We include very large magnetic field values solely to
probe the pattern described above. 

\begin{figure}[htb]
\begin{center}
\centerline{\includegraphics[width=17cm,height=12.0cm]{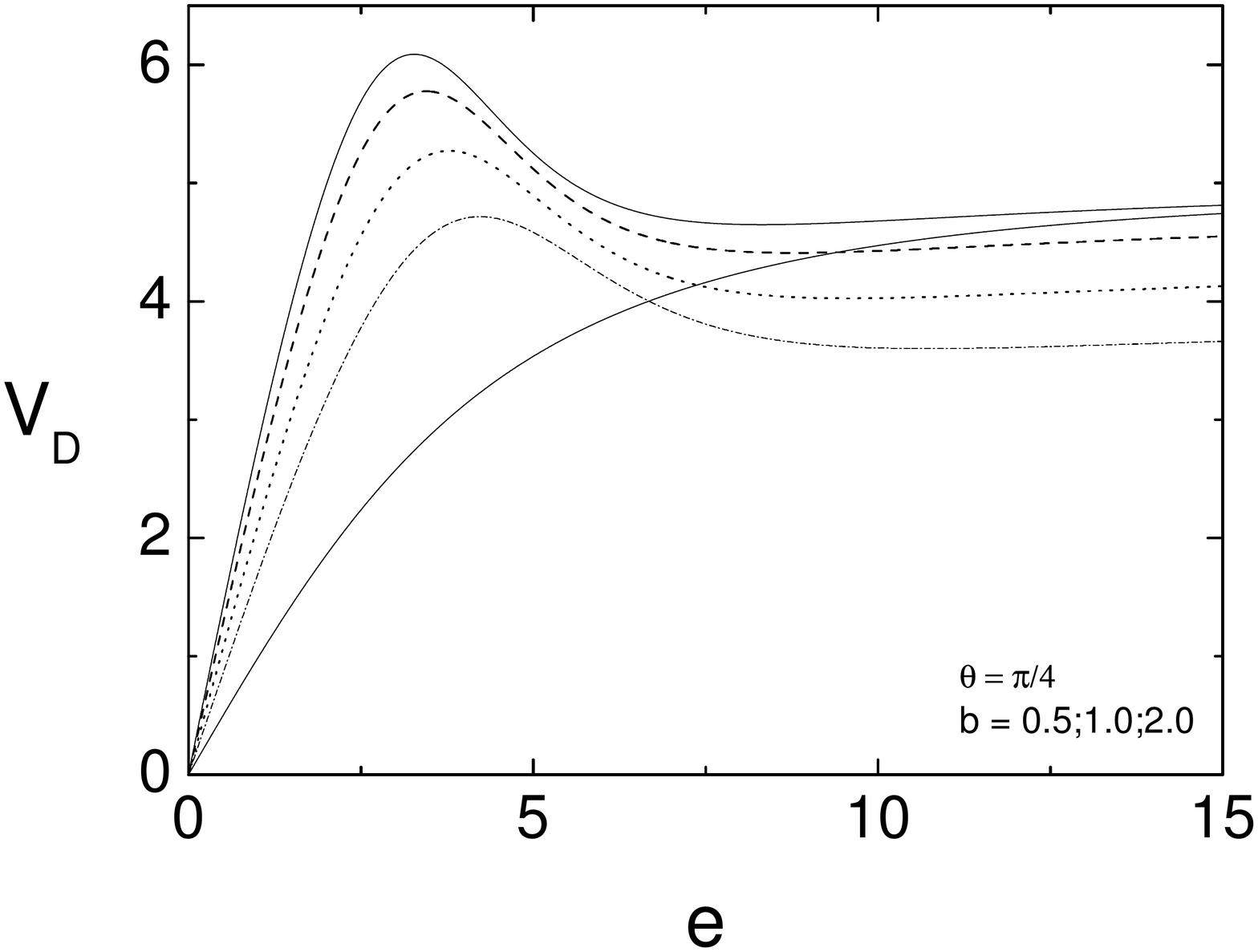}}
\end{center}
\caption{\it{Drift velocity versus electric
field. Upper solid line for zero magnetic filed (or arbitrary magnetic filed
with $\protect\theta =0$). Dashed lines in decreasing order \ for $b=0.5,1.0$
and $2$.$0$, with $\protect\theta =\protect\pi /4.$ Lower solid line case of
null gap and zero magnetic field. See main text for velocity and field units.}}
\label{fig6}
\end{figure}

\begin{center}
\bigskip

\bigskip \newpage
\end{center}

\begin{figure}[htb]
\begin{center}
\centerline{\includegraphics[width=17cm,height=12.0cm]{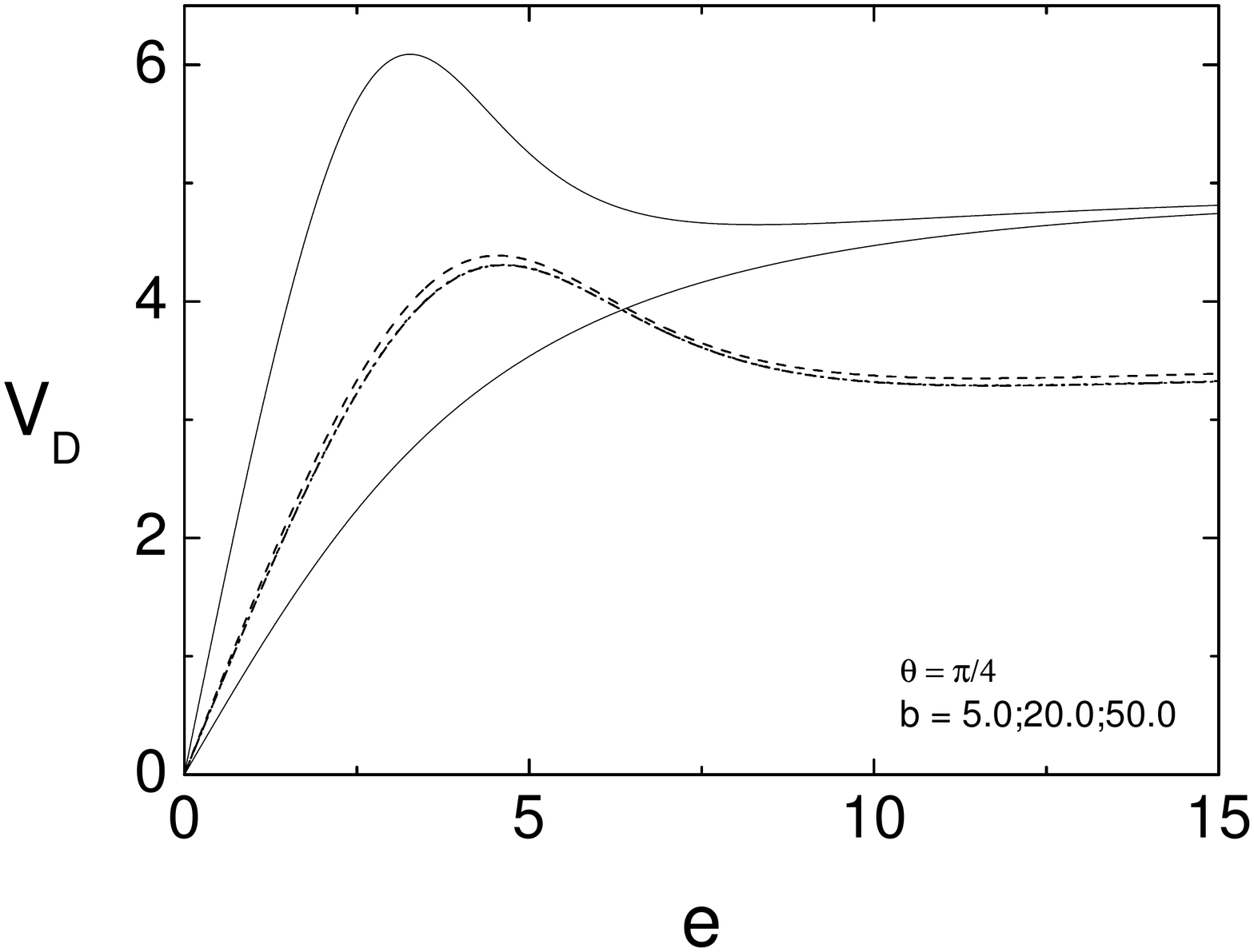}}
\end{center}
\caption{\it{Drift velocity versus electric
field. Upper solid line for zero magnetic filed (or arbitrary magnetic filed
with $\protect\theta =0$). Dashed lines in decreasing order \ for $%
b=5.0,20.0 $ and $50.0$ (the last two cases overlap on the scale used) ,
with $\protect\theta =\protect\pi /4.$ Lower solid line case of null gap and
zero magnetic field. See main text for velocity and field units.}}
\label{fig7}
\end{figure}

\begin{center}
\bigskip

\bigskip \newpage
\end{center}

\begin{figure}[htb]
\begin{center}
\centerline{\includegraphics[width=17cm,height=12.0cm]{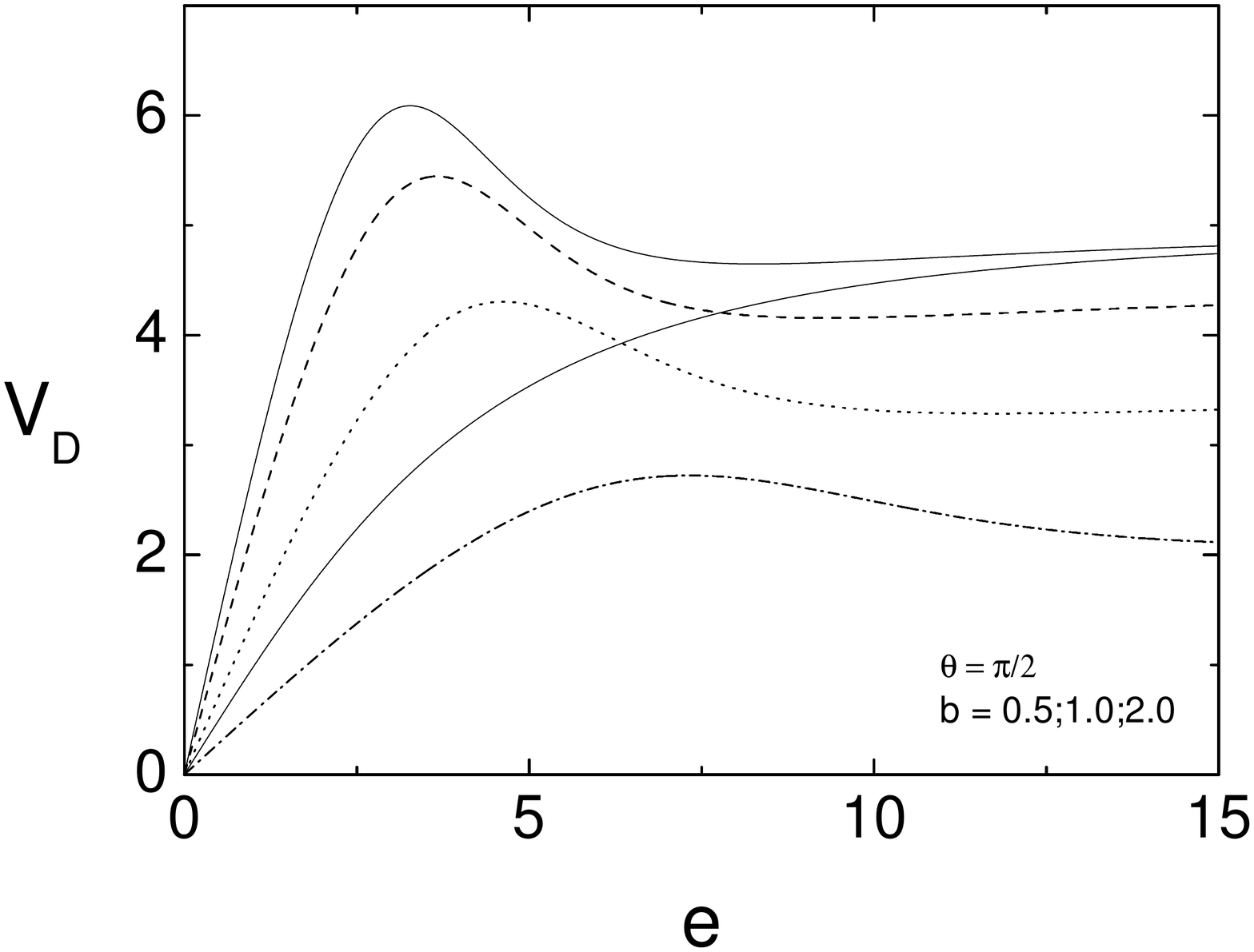}}
\end{center}
\caption{\it{Upper solid line for zero
magnetic filed (or arbitrary magnetic filed with $\protect\theta =0$).
Dashed lines in decreasing order \ for $b=0.5,1.0$ and $2.0$ , with $\protect%
\theta =\protect\pi /2.$ Lower solid line case of null gap and zero magnetic
field. See main text for velocity and field units.}}
\label{fig8}
\end{figure}

\newpage

\begin{figure}[htb]
\begin{center}
\centerline{\includegraphics[width=17cm,height=12.0cm]{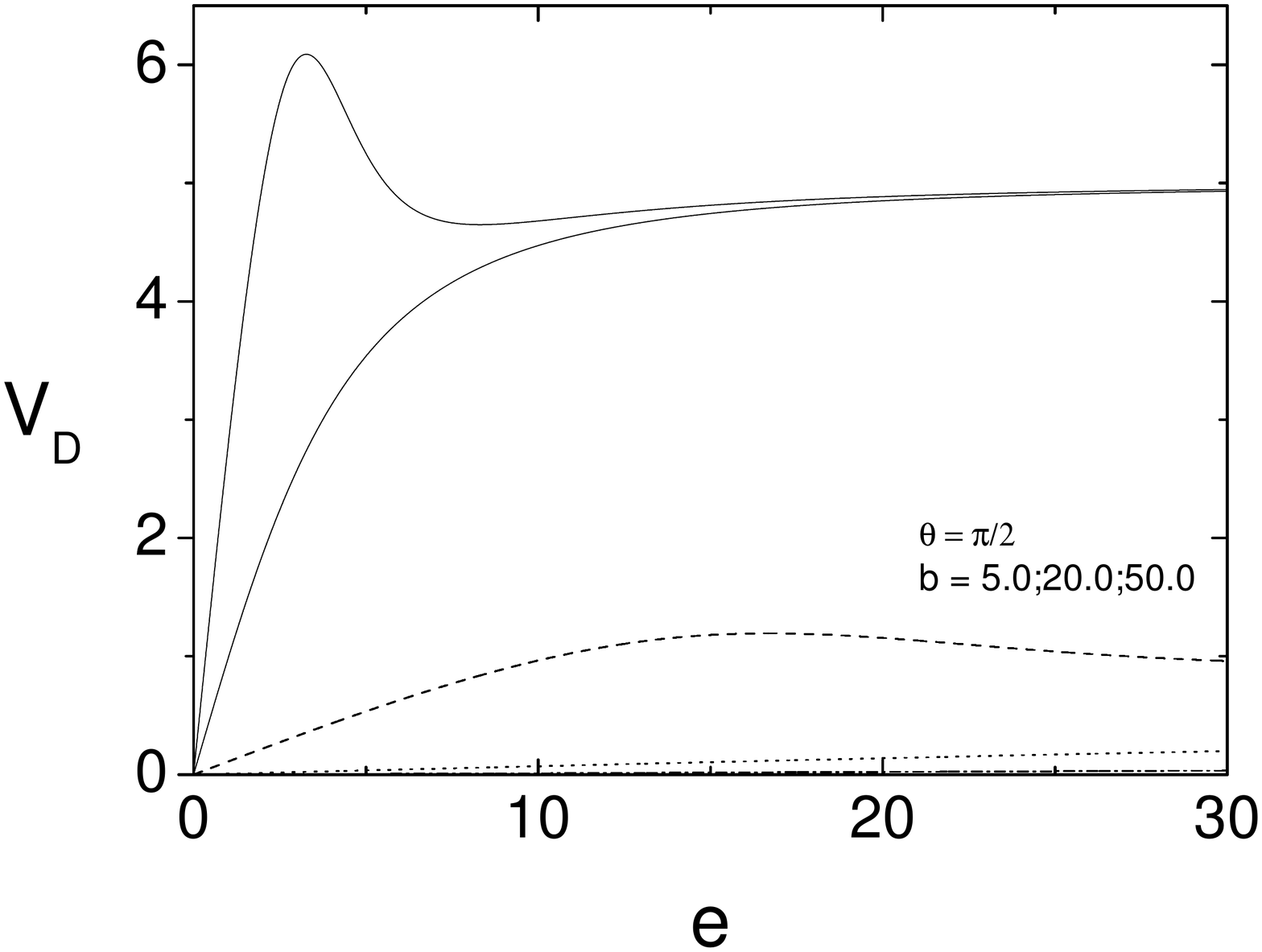}}
\end{center}
\caption{\it{Drift velocity versus electric
field. Upper solid line for zero magnetic filed (or arbitrary magnetic filed
with $\protect\theta =0$). Dashed lines in decreasing order \ for $%
b=5.0,20.0 $ and $50.0$ (the last case almost indistinguishable from the $e$
axis, on the scale used) , with $\protect\theta =\protect\pi /2.$ Lower
solid line case of null gap and zero magnetic field. See main text for
velocity and field units.}}
\label{fig9}
\end{figure}

\begin{center}
\newpage

{\large 5: Concluding remarks }
\end{center}

We presented the Langevin formulation for a Brownian carrier under uniform
and static external electric and magnetic fields. From the solution to the
associated Langevin equation, we computed the relaxation of the carrier's
effective (nonequilibrium) temperature towards a (hot) steady state regime
with a nonequilbrium field dependent temperature. The latter was compared
with well known existing results. We believe our present result in the
Langevin \ as well as our previous results in the Kramers Smoluchowski
scheme, incorporate the effect of the magnetic field hitherto not
considered. Then we presented a simple yet relevant Brownian model to
account for the negative differential conductivity behavior on the GaAs
compound, again incorporating the magnetic field effects hitherto not
considered. Discussions of results and Figures are presented at the end of
sections 3 and 4.

Our future work includes the incorporation of diffusive effects on the
effective temperature, as discussed in section 3, the inclusion of other
scattering mechanisms into the mobility as discussed in section 4, and
incorporate within the Langevin Formalism chemical reactions [61-62] and
photovoltaic effects [63] as discussed for example in [45] \ within the
Kramers Smoluchowski context. 

As a final remark, we comment on the several techniques employed to solve 
the Brownian Motion Problem in Fields of Forces, following from Kramers original
mathematical acrobacies [9]. We mention Chandrasekhar's proposal 
of six independent first integrals within a Gaussian ansatz [10]; tensorial 
frictional forces [25]; gauge transformations [27]; a combination of several 
of the above mentioned techniques [28,45] and the time-dependent rotation 
matrix method [29-31,36-39,44]. Here, in this paper (as in [47]) we 
directly apply the Cayley-Hamilton theorem. Paraphrasing Professor R. Kubo
from his opening address [64]: If we borrow the terms from quantum mechanics,
the Fokker-Planck (Kramers, Smoluchowski) equation is the Schroedinger picture
and the Langevin equation is the Heisenberg picture of the same problem. 
One can go from one to another, allowing for intermediate (mixed) representations. 

In this context, we may regard as equivalent all the techniques mentioned above,
when pursuing the exact solution of this linear problem. All these methods exhibit 
advantages and disadvantages when compared to each other, depending on the starting
point, namely the Fokker-Planck or the Langevin representation; on the particular
physical quantities to be computed or the particular regimes to be studied 
(homogeneous fields, overdamped and/or inertial regime et cetera).
\ 

\noindent \emph{Acknowledgments}: \ We thank Profs. Ricardo Paupitz and
Mariano de Souza for \ valuable comments. This work was partially supported
by CNPq (Brasil).

\end{document}